\documentclass[12pt,a4paper]{article}
\usepackage{latexsym,amssymb,amsmath,epsfig,graphicx}
\usepackage{pstricks,color,amscd}
\newcommand{\pkt}{\; .}
\newcommand{\kma}{\; ,}

\newcommand{\beast}{\begin{eqnarray*}}
\newcommand{\eeast}{\end{eqnarray*}}

\newcommand{\cala}{{\cal A}}
\newcommand{\calp}{{\cal P}}

\newcommand{\tr}{{\rm tr}}

\newcommand{\be}{\begin{eqnarray}}
\newcommand{\ee}{\end{eqnarray}}

\begin{document}
\title
{Inhomogeneous baryogenesis, cosmic antimatter, and dark matter}
\author{A.D.~Dolgov$^{\rm 1,2,3,4}$,
M. Kawasaki$^{\rm 4}$, N.~Kevlishvili$^{\rm 1,2,5}$
\\[3mm]
$^{\rm 1}$
Universit\`a degli Studi di Ferrara, I-44100 Ferrara, Italy \\
$^{\rm 2}$INFN, 
Sezione di Ferrara, I-44100 Ferrara, Italy \\
$^{\rm 3}$ITEP,
113259 Moscow, Russia \\
$^{\rm 4}$ ICRR, Tokyo, Japan\\
$^{\rm 5}$ Andronikashvili Institute of Physics, GAS, 0177 Tbilisi, Georgia
}
\maketitle


\begin{abstract}
A model of inhomogeneous baryogenesis based on the Affleck and Dine mechanism is described. 
A simple coupling of the scalar baryon field to the inflaton allows for formation of 
astronomically significant bubbles with a large baryon (or antibaryon) asymmetry.
During the farther evolution these domains form compact stellar-like objects, or lower 
density clouds, or primordial black holes of different size. According to the scenario, 
such high baryonic number objects occupy relatively small fraction of space but despite
that they may significantly contribute to the cosmological mass density. For some values 
of parameters the model allows the possibility the whole dark matter in the universe to be 
baryonic. Furthermore, the model allows the existence of the antibaryonic B-bubbles, i.e. 
a significant fraction of the mass density in the universe can be in the form of the compact 
antimatter objects (e.g. anti-stars). 
\end{abstract}

\section{Introduction \label{s-intr}} 

Though the universe in our neighbourhood looks as 100\% baryonic, the
observations do not exclude subdominant but astronomically large domains and 
objects of antimatter. For a recent analysis and literature see e.g.  
ref.~\cite{Casadei:2004pi,Bambi:2007cc}. The standard mechanism of baryogenesis~\cite{Sakharov:1967dj}
predicts baryon asymmetry expressed as the ratio of baryon-to-photon
number densities:
\be
\beta = n_B/n_\gamma
\label{beta}
\ee
to be constant both in space and, after baryogenesis is over, also in time, 
however, more complicated models can lead to not so dull pattern. The outcome of
the baryogenesis scenario is closely related to the mechanism of CP-violation realized 
in cosmology. For a review of the latter see ref.~\cite{Dolgov:2005wf}.  
If CP is broken spontaneously~\cite{Lee:1974jb}, the universe would be charge symmetric
with equal weight of baryonic and antibaryonic domains so that the average value of
$\beta$ over whole universe would be zero~\cite{Zeldovich:1974uw,Brown:1979dq}. 
However, in the standard cosmology the domain size would be
very small by astronomical standards and some exponential (inflationary) expansion is 
necessary to avoid immediate problems~\cite{Sato:1980yn,Sato:1981ds}. 
Another cosmological problem of domain walls with huge energy density inside our 
horizon~\cite{Zeldovich:1974uw} can be solved
either by non-restoration of CP invariance at high 
temperature~\cite{Mohapatra:1979vr,Mohapatra:1979qt}, but in this
case antimatter domains would not be created, or by a possible mechanism of wall 
destruction~\cite{Kuzmin:1981ip,Kuzmin:1981xu}. However, even if these problems
are resolved, it can be shown that baryo-symmetric cosmology demands the domain size above
(a few) Gpc. Otherwise too 
intensive gamma ray background would be created by the annihilation on the domain boundaries
~\cite{Cohen:1997ac}.

If in addition to spontaneous CP-violation there exists an explicit one, the universe
would be baryo-asymmetric and, if the magnitude of spontaneous CP breaking happened to be
larger than the explicit one, there could exist subdominant antimatter domains
~\cite{Kuzmin:1981ip,Kuzmin:1981xu,Chechetkin:1982xy}.

Here we explore another mechanism of inhomogeneous baryogenesis
suggested in ref.~\cite{Dolgov:1992pu}. This mechanism is based on the  
Affleck and Dine~\cite{Affleck:1984fy} idea that in supersymmetric theories a 
complex scalar field $\chi$
with non-zero baryonic number may be displaced far from the origin,
$\chi=0$, along flat directions of the potential $U(\chi)$. Such flat 
directions are generic features of supersymmetric models. 
Inflationary stage is especially favorable for such a travel along the
flat directions due to the 
infrared instability of massless or light, $m^2_\chi \ll H^2$,
fields in De Sitter space-time. According to the results of ref.~\cite{Linde:1982uu, Vilenkin:1982wt, Starobinsky:1982ee} 
(see however, ref.~\cite{Dolgov:2005se}), vacuum expectation value of massless scalar field 
rises as:
\be
\langle \chi^2 \rangle = H^3 t /4 \pi^2\pkt
\label{chi-2-vac}
\ee
Inflation not only drives $\chi$ away from the origin but also allows for
formation of the exponentially big domains with a large value of $\chi$.

After the Hubble parameter, $H$, dropped below $ m_\chi$, the scalar baryon 
(AD-field) $\chi$
would evolve down to $\chi = 0$ transforming the accumulated baryonic charge
into quarks by $B$-conserving decays. Typically such scenario leads to a large
baryon asymmetry, $\beta \sim 1$ and theoretical efforts should be taken to 
diminish it down to the observed value $\beta \approx 5\cdot 10^{-10} $.

The modification of this mechanism, according to ref.~\cite{Dolgov:1992pu}, was an
introduction of the quadratic coupling of $\chi$ to the inflaton field $\Phi$ in 
the following renormalizable form, $\lambda_1 |\chi|^2 (\Phi^2 +\tilde{\mu} \Phi )$,
where $\tilde{\mu}$ is a constant parameter with dimension of mass.
This term can be rewritten as:
\be
U_{int} (\chi,\Phi) = \lambda_1 |\chi|^2 \left( \Phi - \Phi_1\right)^2
\label{U-int}
\ee
by redefinition of the mass term of $\chi$. 

When the inflaton field evolves down to zero, it passes through $\Phi=\Phi_1$
and at that moment the window to flat direction in $\chi$ potential would be
open but only for a rather short period of time. Correspondingly, the probability
for $\chi$ to evolve away from the origin would be small and baryogenesis
resulting in high $\beta$ could proceed only in a minor fraction of space. 
So we should expect that in the bulk of space the usual homogeneous baryogenesis
took place, and compact B-bubbles, with very high baryonic number density
could be created over this background. Despite their small size, B-bubbles might
contribute significant amount into the total cosmological mass density and even make
all dark matter of the universe. They would either form compact stellar-like
objects, or high density clouds, or primordial black holes (BH). In the simplest
version of the model there should be an equal number of baryonic and antibaryonic
B-bubbles. So cosmology would be almost baryo-symmetric but the bounds
derived in ref.~\cite{Cohen:1997ac} are not applicable here because of compactness of 
the bubbles. An interesting feature of this scenario is that dark matter (DM)
can be made from the usual particles, baryons and antibaryons in the form of evolved
stellar type objects or primordial black holes.

The paper is organized as follows. In the Sec. \ref{sec-model} we give a detailed 
description of the model and it's parameter space. The B-bubble evolution
and the generation of the baryon asymmetry are discussed in Sec. \ref{s-b-evol}.
The next section, Sec. \ref{s-size}, gives a detailed description of the
development of quantum fluctuations of the Affleck and Dine field and the distribution 
of B-bubbles over the length and mass. In Sec. \ref{s-perturb} the fate of B-bubbles 
during different cosmological epochs is considered. In Sec.\ref{s-cosm-astro} the 
observational bounds on the dark matter in the form of primordial black holes (PBH) and 
the amount af antimatter in the compact stellar objects are presented.
Our conclusions are hosted in the Sec. \ref{s-conclusions}.

\section{The model and the parameter space}\label{sec-model}
First we will study qualitatively the evolution of $\chi$ and formation of the 
bubbles with a large value of this field. Though the model and results are similar to 
those of ref.~\cite{Dolgov:1992pu}, here we discuss a somewhat simpler
scenario.

Let us consider a toy model which can lead to the features
described in the Introduction. We assume that the evolution of the
inflaton field is governed by the usual potential:
\be
U_\Phi(\Phi) = m_\Phi^2 \Phi^2/2 + \lambda_\Phi \Phi^4/4,
\label{U-of-Phi}
\ee
while $\chi$ ``lives'' in the potential 
\be
U_\chi (\chi, \Phi) = \lambda_1(\Phi-\Phi_1)^2|\chi|^2
+\lambda_2|\chi|^4\ln{\frac{|\chi|^2}{\sigma^2}}+ m_0^2 |\chi|^2
+ m_1^2\chi^2+m_1^{*2}\chi^{*2}.
\label{U-of-chi-Phi}
\ee
The first term in this expression describes the interaction between
the inflaton and $\chi$ fields introduced in ref.~\cite{Dolgov:1992pu}. We
assume that $\lambda_1$ and $|\chi|$ are sufficiently small, and so 
this term has negligible impact on the inflaton evolution.
The second term is the Coleman-Weinberg potential~\cite{Coleman:1973jx} obtained by 
summation of one loop corrections to the quartic potential, $\lambda_2 |\chi|^4$.
The last two mass terms are not invariant with respect to the phase
rotation:
\be 
\chi \rightarrow e^{i\theta} \chi
\label{chi-transform}
\ee
and thus break baryonic charge conservation (see below). If 
$m_0^2 < 2|m_1|^2$, the potential created
by these two terms is not bounded from below but the effective mass:
\be
m_{eff}^2=\lambda_1(\Phi-\Phi_1)^2+ m_0^2 +
2|m_1|^2 \cos (2\alpha + 2\theta)
\label{m-eff}
\ee 
is almost always positive, except for a short time when $\Phi \approx \Phi_1$.
Anyhow the $\lambda_2|\chi|^4 \ln |\chi|^2$ term stabilizes the potential but
its minimum shifts from $\chi^{(1)}=0$ to some non-zero $\chi$. In the above
equation $\alpha$ and $\theta$ are the phases of mass, $m_1=|m_1|\exp (i\alpha)$, 
and of AD-field, $\chi =|\chi| \exp (i\theta)$, respectively.

It is interesting that potential (\ref{U-of-chi-Phi}) conserves CP despite
presence of the complex mass parameters. Indeed the imaginary part of $m$ 
can be excluded by the redefinition $\chi\rightarrow \chi \exp (-i\alpha)$.
However, a coupling of $\chi$ to fermions may destroy CP-conservation.

Classical evolution of homogeneous fields $\Phi$ and $\chi$ is determined by the 
equations
\be
\ddot \Phi + 3 H\Phi + m_\Phi^2 \Phi + \lambda_\Phi \Phi^3 +
2\lambda_1 |\chi|^2 (\Phi - \Phi_1) = 0\kma  
\label{ddot-Phi}
\ee
\be
\ddot \chi + 3H \dot\chi + \lambda_1 (\Phi-\Phi_1)^2 \chi + 
\lambda_2 |\chi|^2 \chi \left(2 \ln \frac{|\chi|^2}{\sigma^2} +1\right) 
+ 2m^{*2}\chi^* = 0\kma
\label{ddot-chi}
\ee
which we solved numerically, but qualitative behavior of the solutions can be
understood from the following simple considerations. We start from inflationary
stage when the cosmological energy density is dominated by the inflaton field. 
We assume that the product $\lambda_1 |\chi|^2$ is sufficiently small, so the
$\chi$-dependent term in eq. (\ref{ddot-Phi}) is not essential. In this regime
the evolution of $\Phi$ is simply a slow roll in the corresponding 
potential~(\ref{U-of-Phi}) till the Hubble parameter remains large in comparison
with the potential slope:
\be
H^2 = \frac{8\pi}{3m_{Pl}^2}\,\left( \frac{m_\Phi^2 \Phi^2}{2} + 
\frac{\lambda_\phi \Phi^4}{4}\right) \gg U''(\Phi)=m_\Phi^2 + 3\lambda_\Phi \Phi^2\pkt
\label{H2}
\ee
In the course of the cosmological expansion $\Phi$ approaches the equilibrium point
$\Phi =0$ and $H$ drops down so that 
\be
H^2 \leq m^2 (\Phi) = m_\Phi^2 + 3\lambda_\Phi \Phi^2\pkt
\label{m-of-Phi}
\ee
Inflation stops and $\Phi$ starts to oscillate around $\Phi = 0$ producing particles and
heating the universe. Correspondingly, the De Sitter expansion regime with domination of 
the vacuum-like energy turns into the matter domination and later into the radiation domination.
To be more precise the equilibrium value of $\Phi$ depends upon $\chi$. But as we will see 
below, $\chi$ ultimately tends to zero and during inflationary period
this correction is not of much importance. If the life-time of $\chi$ is longer
than the life-time of the inflaton, the impact of $\chi$ on the inflaton
behavior might be essential at the (re/pre)heating.

According to our assumption and with our choice of the parameters, 
the impact of $\chi$ on the inflaton behavior is almost unimportant. Therefore during the
qualitative discussion we consider the evolution of $\chi$
in the fixed (but still time dependent) inflaton background. 
However, we have not neglected corresponding term when solved the equations numerically.

At the beginning of inflation, $\Phi$ is large, $\Phi > m_{Pl}$, and the effective
mass of $\chi$ (\ref{m-eff}) is also large. During this period potential 
(\ref{U-of-chi-Phi}) has only one minimum at $\chi = 0$ and if $m_{eff}$, 
eq. (\ref{m-eff}), is large in comparison with $H$, quantum fluctuations of 
$\chi$ are not essential and field $\chi$ ``sits'' in the minimum $\chi^{(1)} = 0$. 

In fact the picture is more complicated. If $\lambda_\phi \neq 0$, then
for large $\Phi$ the Hubble parameter would be larger than $m_{eff}$ and
$\chi$ could quantum fluctuate away from zero climbing up the potential
slope during an early inflationary stage. With decreasing $H$ field $\chi$
would return back to zero if the classical rolling down of $\chi$ is faster
than the decrease of $H$.

\begin{figure}
	\centering
		\includegraphics[scale=0.6]{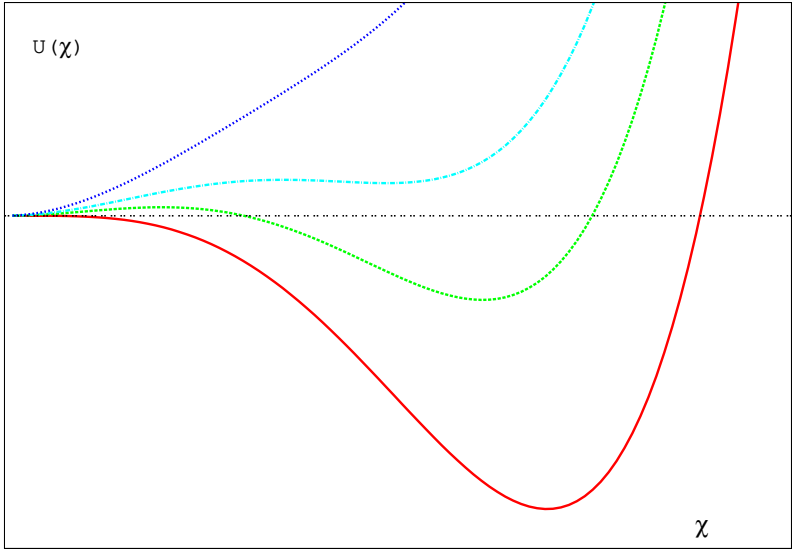}
		\caption{\small Qualitative behaviour of $U_\chi(\chi)$ for different values of
    $m^2_{eff} (t)$.}
	\label{fig:Potevolution}
\end{figure}

When the amplitude of the inflaton drops down and approaches $\Phi_1$, 
the effective mass of $\chi$ decreases and the potential gets the second minimum at 
some $\chi^{(2)}\neq 0$. This second minimum appears and disappears
when $m^2_{eff}$  passes down and up the value $2\lambda_2 \sigma^2 /e^{3/2}$.
Initially the second minimum at $\chi = \chi^{(2)}$
lays above the minimum at $\chi^{(1)}=0$, and later with the diminishing difference
$|\Phi - \Phi_1|$ it becomes the absolute minimum, but separated from the
minimum at $\chi^{(1)}=0$ by a barrier. 
In this situation the first order phase transition by quantum tunneling from
$\chi =0$ to $\chi = \chi^{(2)}$ is possible, but the transition probability 
is exponentially small and the process can be neglected.
The transition to $\chi =\chi^{(2)}$ would be much more efficient when the
barrier became low or disappeared. It is essential that the low barrier
existed only for some finite time. After $\Phi$ passed $\Phi_1$ and traveled
further to zero, the potential barrier would reappear. 
The behavior of $U(\chi,\Phi)$ is depicted in Fig. \ref{fig:Potevolution}.
As a result of such non-monotonic behavior of the potential, $\chi$ would be
either forced back to the first minimum at $\chi =0$ or would remain beyond 
the barrier near $\chi=\chi^{(2)}$. The separation depends upon the amplitude
which $\chi$ managed to reach to the moment of the ``gate closing''.
The boundary/critical value of $\chi$ is determined by the position of the 
maximum of $U(\chi,\Phi)$ at $\chi = \chi^{(3)}$, which separates the two minima 
(see Fig. \ref{fig:Potevolution}). $\chi^{(3)}$ is the smaller solution of the  
equation:
\be
2\lambda_2 |\chi|^2 \,\ln\frac{\sqrt{e} |\chi|^2}{\sigma^2}
+m^2_{eff} = 0
\label{eq-chi-3}
\ee
and by an order of magnitude: $\chi^{(3)}\sim m_{eff}/\sqrt{\lambda_2}$. An important
feature of $\chi_3$ is that it is not constant but moves together with
$\Phi$.

\normalsize

Thus we expect that for some time, when the second minimum at $\chi^{(2)}$
existed, there should be two phases with large $\chi \sim \chi^{(2)}$ and small 
$\chi \sim 0$. Which phase dominates is determined by the duration of the 
``open gate'' period and the efficiency of quantum fluctuations
of $\chi$. We will show in what follows (Sec. \ref{s-size})
that one can naturally realize the
scenario when the probability of formation of large $\chi$ bubbles is sufficiently
small and they would occupy a small fraction of the total volume of the universe.
Still their size can be astronomically large if they were formed during inflationary
stage. In other words, when $\Phi$ was close to $\Phi_1$, inflation lasted
sufficiently long to stretch up the bubble size exponentially. In the future such 
bubbles would form objects (or domains) with a large value of the baryon 
asymmetry and we will call them B-bubbles or just BBs.

Note that the phase transition (p.t.) which we discuss here is neither first order,
nor second order but something in between. It takes place when the barrier is
absent and in this sense it is second order but it resulted in formation of
bubbles of new phase inside the old one and, as such, it resembles the first 
order p.t. Note that the evolution of $\chi$ demonstrates the
hysteresis behavior. Initially when $\chi$ was near zero, it remained there
despite appearance of the second minimum of the potential at $\chi^{(2)}$. 
At a later stage when the second minimum became higher than the first one,
$\chi$ remained there and rolled down to $\chi=0$ only when the minimum
$\chi^{(2)}$ completely disappeared.

We assume that $m_0^2 > 2 m_1^2$, so the local minimum of $U_\chi$ never disappears
and classically $\chi$ should always remain zero. However, quantum fluctuations of
light, $m<H$, scalars in De Sitter background allow to overcome this barrier.
Roughly speaking it is possible when $\lambda_1 (\Phi - \Phi_1)^2<H^2$ and
thus the characteristic time when the gate to the negative slope of
the potential was open, lasted approximately
\be
\Delta t \approx \frac{3H^2}{\sqrt{\lambda_1} U'_\Phi} =
\frac{2\pi \Phi_1 (2 m_\Phi^2 +\lambda_\Phi \Phi_1^2)}
{\sqrt{\lambda_1} m_{Pl}^2  ( m_\Phi^2 +\lambda_\Phi \Phi_1^2)}\kma
\label{Delta-t}
\ee
where we used the slow roll approximation: $\dot\Phi \approx U'_\Phi/3H$.
According to eq. (\ref{chi-2-vac}), during this time interval $\chi$ 
can reach the value of the order of
\be
\chi^2 \sim \frac{H^3\Delta t}{4\pi^2} \sim 
\sqrt{\frac{2\pi}{27\lambda_1}}\,\frac{\Phi_1^4}{m_{Pl}^5}\,
\frac{(2 m_\Phi^2 +\lambda_\Phi\Phi_1^2)^{5/2}}
{ m_\Phi^2 +\lambda_\Phi\Phi_1^2}
\label{chi-2}
\ee
For the chosen below parameters $\chi$ can reach the boundary value $\chi_c$ after which it would 
evolve classically down to the second minimum of $U_\chi$.

Since during this period $m_1^2$ was much smaller than $H^2$,
the phase of $\chi$ equally populated
the interval $[0,2\pi]$. This would allow for generation of 
bubbles with a large baryon asymmetry. On the other hand, if $\chi$ stays in 
the minimum over the phase variable i.e. at 
$\cos(2\alpha + 2\theta) =-1$, the baryon asymmetry would be small, because
no ``angular momentum'' can be generated in the process of the evolution of
$\chi$ down to zero, despite the non-sphericity of the potential 
(see Sec.~\ref{s-b-evol}).
Such a case could be realized if $\chi$ evolved to $\chi_2$ along the
path $\alpha+\theta = \pi/2 $. However, even in this case the $\chi$-state
initially localized near this value of the phase, would be dispersed by quantum 
fluctuations over all values of the phase because the mass of the 
pseudogoldstone degree of freedom is much smaller than the Hubble
parameter.

In flat space-time the tunneling from $\chi^{(1)}=0$ to a nonzero $\chi$ 
in the potential $\lambda |\chi|^4$ is determined by the Fubini 
instanton~\cite{Fubini:1976jm},
which leads to a very low probability of the transition, 
$\sim \exp(-32\pi^2/3\lambda)$. The exponential suppression
for small $\lambda$ is related to a non-zero energy density of the bubble 
wall, where $\chi$ interpolated from $\chi =0$ to $\chi = \chi^{(2)}$.  
This suppression is absent
in our case by the following reasons. First, the effective coupling
strength, $\lambda_2\, \ln (|\chi|^2/\sigma^2)$, in our case is not constant but
diverges at $\chi = 0$. Second, the exponential expansion stimulates rise
of quantum fluctuations and leads to a large deviations from $\chi =0$.
Moreover, all space inhomogeneous effects are exponentially smoothed down by
the cosmological expansion. A detailed discussion of tunneling in 
curved space-time can be found in ref.~\cite{Goncharov:1986ua}

Another danger comes from the opposite direction. The probability of
BB formation may be too large and practically all the space would be
occupied by the large $\chi$ bubbles, while low $\chi$ values would be 
practically absent. However, this problem can be easily cured by an
adjustment of the parameters of the model, i.e. of the masses and coupling
constants in potentials (\ref{U-of-Phi}) and (\ref{U-of-chi-Phi}). 
A sufficiently large term $m_0^2 |\chi|^2$ can lead to the necessary suppression
of the bubble formation.

The picture that we described above can be realized with the following
rather natural values of the parameters. Since we want inflation to 
continue when $\Phi$ passes $\Phi_1$, we need $\Phi_1>m_{Pl}$; 
$\Phi_1 = (2-3) m_{Pl}$ is sufficient for creation of astronomically large B-bubbles.
The inflaton
mass should be bounded by $10^{-7}m_{Pl}\lesssim m_\Phi\lesssim 10^{-6}m_{Pl}$ 
to avoid too large density perturbations, $\delta \rho /\rho < 10^{-5}$
~\cite{1990ppic.book.....L}. The coupling constant $\lambda_\Phi$ must be in the interval
$10^{-14}\lesssim\lambda_\Phi\lesssim10^{-12}$ by the same reason to 
suppress the density perturbations which at horizon crossing are, roughly 
speaking, equal to $\delta \rho /\rho \sim 10^2\sqrt{\lambda_\Phi}$~\cite{1990ppic.book.....L}.
The coupling of $\chi$ to the inflaton, $\lambda_1$, is bounded by
$\lambda_1<10^{-6}$, because otherwise the one loop correction to the inflaton
potential induced by this coupling would give rise to unacceptably large
$\lambda_\Phi \sim \lambda^2_1/4\pi^2$. Interesting results are obtained with
$\lambda_1 \sim 10^{-10}$. Such a small value of the inflaton coupling to
$\chi$ naturally fits the idea of very weak coupling of the inflaton to the
usual matter.
We choose the $\chi$-self-coupling constant 
$\lambda_2$ to be near $10^{-2}$ to obtain cosmologically interesting consequences.

\section{B-bubble evolution and  baryogenesis in the very early universe}
\label{s-b-evol}

When $\Phi$ dropped noticeably below $\Phi_1$, the potential barrier reappeared,
the formation of the bubbles with $\chi\sim\chi^{(2)}$  from $\chi = 0$ 
terminated, and the primeval cosmological matter would consist of the
inflaton field, which still continued to inflate the universe, with 
exponentially expanding bubbles of non-zero $\chi$ over practically 
homogeneous background with $\chi =0$. The energy density of BBs were initially 
negligible with respect to the total cosmological energy density, 
dominated by $\Phi$, and in a good approximation
the universe can be considered as homogeneous.

Further decrease of the inflaton field leads to a large rise of the effective
mass of $\chi$, eq. (\ref{m-eff}), so the second minimum of the potential 
$U(\chi,\Phi)$ disappears (see Fig. \ref{fig:Potevolution}) and $\chi$ started 
to roll down to zero. The evolution of the spatially homogeneous field $\chi=\chi(t)$
and generation of baryon asymmetry 
can be easily visualized with the following mechanical analogy.
The equation of motion for $\chi$ in arbitrary potential reads:
\begin{equation}
\ddot{\chi}+3H\dot{\chi}+U'(\chi)=0\pkt
\label{ddot-chi-2}
\end{equation}
It is exactly the equation of motion of a point-like particle in potential $U(\chi)$
in Newtonian mechanics. The second term originating from the 
cosmological expansion corresponds to the liquid friction force. In this
language the baryonic charge density is simply
the angular momentum of the particle
\begin{equation}
j_\mu^B=i(\chi^*\partial_\mu\chi-\partial_\mu\chi^*\chi)
=-2|\chi|^2\partial_\mu\theta\pkt
\label{j-B}
\end{equation}
If the potential 
is rotationally symmetric in the complex $\chi$ plane, the angular momentum is 
evidently conserved. 
This is the reason for the introduction of the rotation non-invariant mass
terms in the potential: they break the baryonic current conservation:
\begin{equation}
\partial^\mu j_\mu^B=
2i(m_1^{*2}\chi^{*2}-m_1^2\chi^2)=2|\chi|^2|m_1|^2\sin{2(\theta+\alpha)}\neq 0,
\label{dj}
\end{equation}
where $\theta$ and $\alpha$ are the phases of $\chi$ and $m_1$, respectively.

\begin{figure}[htbp]
 \hspace{-1cm}
 \parbox{7cm}{
		\includegraphics[scale=0.45]{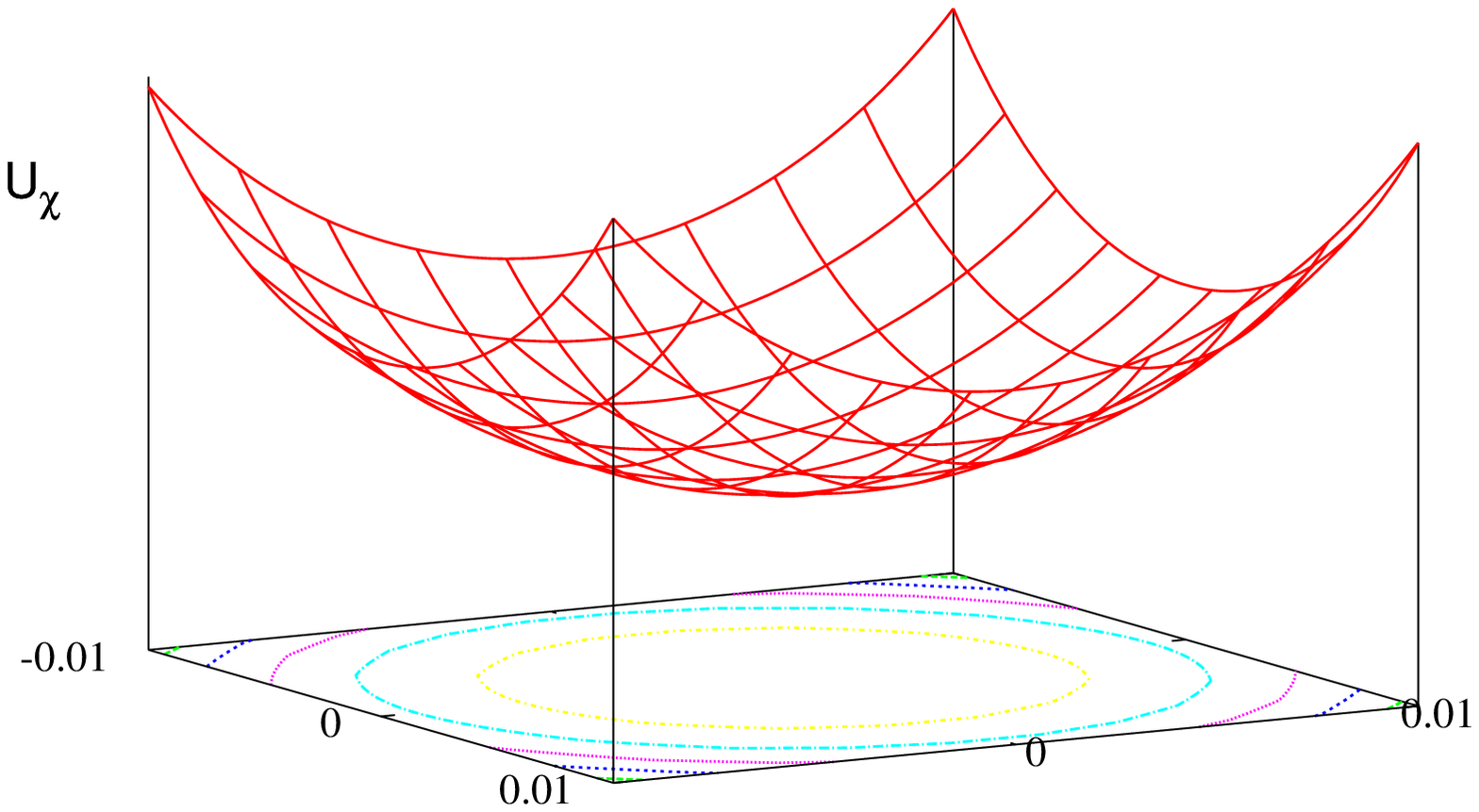}}
 \parbox{5cm}{
    \includegraphics[scale=0.45]{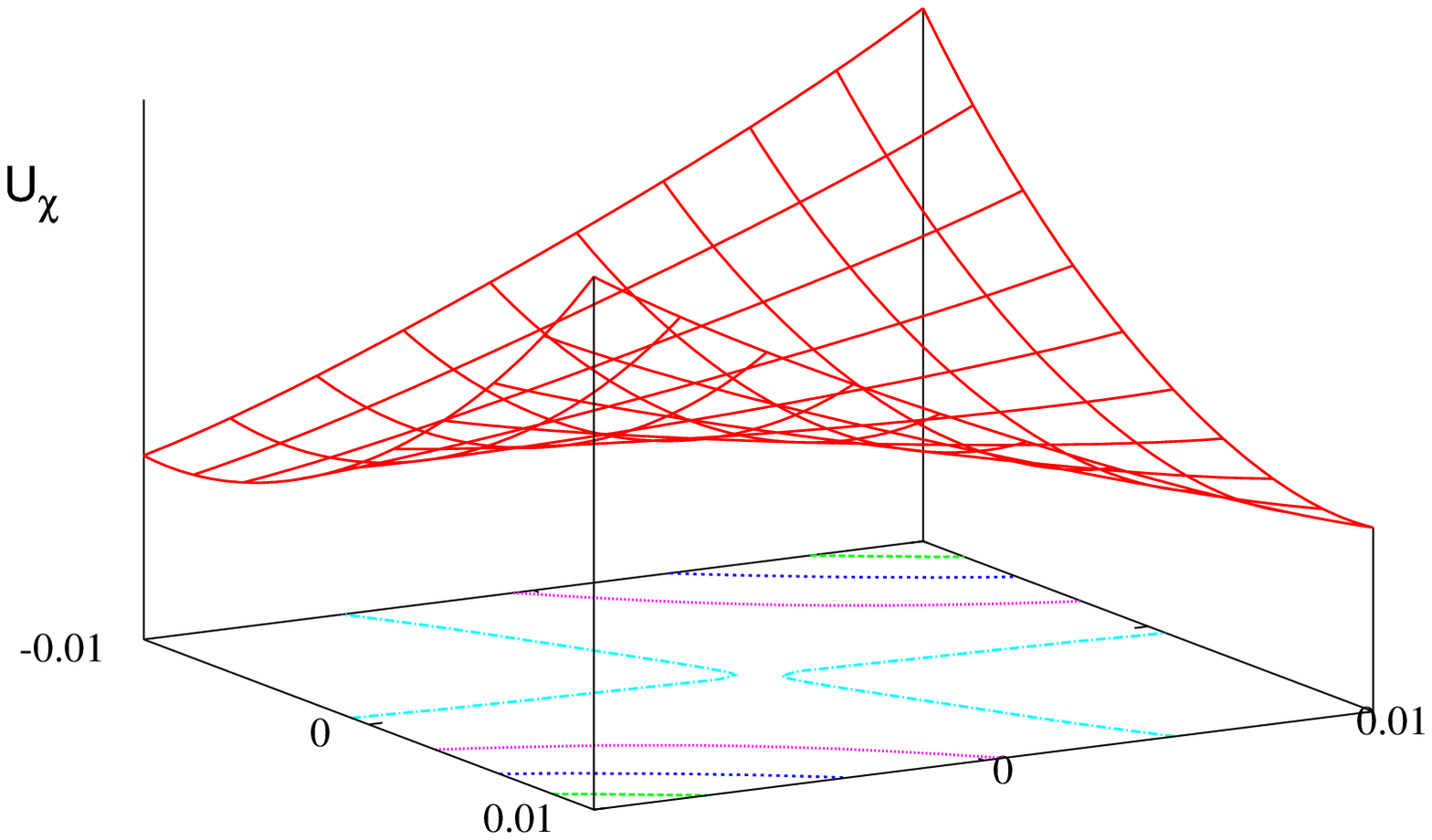}}
    \caption{The form of $U_\chi$ in the complex $\chi$-plane near
the origin. On the left:  $m_1$
    is small and the potential is almast symmetric. On the right: $m_1$ is big and this
    changes the form of the potential, the minimum becomes a saddle point.}
	\label{fig:Potsymasym}
\end{figure}

Too large values of $m_1$ may change the essential features of the potential
described in the previous section, as it is shown on the Fig.~\ref{fig:Potsymasym}. 
In particular, the negative term in $m^2_{eff}$ (\ref{m-eff}) should be smaller than $\lambda_1 \Phi_1^2$
to forbid traveling of $\chi$ from zero to larger values when $\Phi$ is far
from $\Phi_1$. A stronger bound on $m_1$ follows from the requirement that 
the classical slow roll along the minimum at $2(\alpha+\theta)=\pi$, is not
too fast. According to eq. (\ref{ddot-chi}) this would be avoided if $m_1$ is bounded 
from above by $|m_1|<m_\Phi$. The classical roll down from $\chi =0$ would be 
absent for any $\Phi$ if $m_0^2 >2|m_1|^2$

\begin{figure}[htbp]
\vspace*{16mm}
	\centering
	\hspace*{-7cm}
		\includegraphics[scale=0.6]{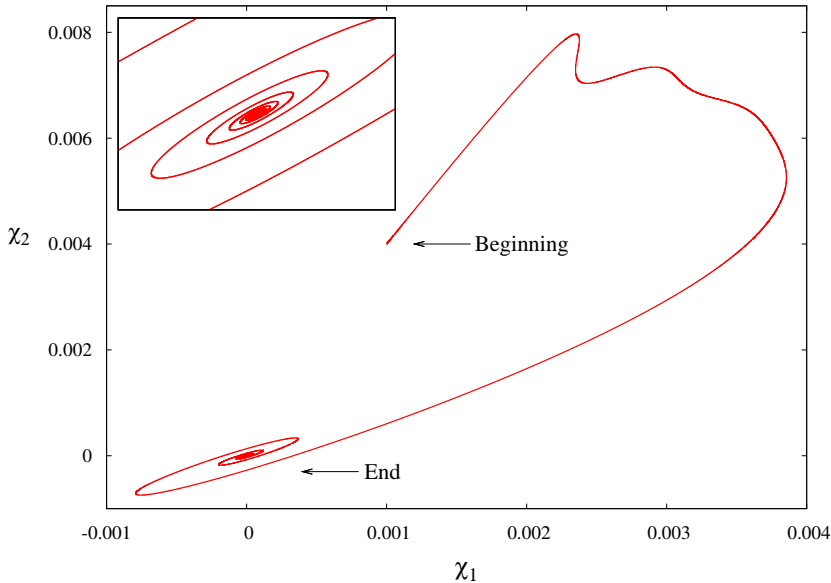}
		\caption{The typical picture of the evolution of $\chi$  
		in the complex $\chi$-plane, $\chi= \chi_1 +i\chi_2$. The
zoomed part inside the figure shows the rotation around
		the absolute minimum after the end of inflation.}
	\label{fig:Chievol}
\end{figure}

For such values of the parameters the non-sphericity of the potential is quite
small and there is no preferred direction in the complex $\chi$-plane for the 
quantum fluctuations generated at inflationary stage when $m_{eff}<H$. On the other hand,
when inflation is over and $\chi$ approaches its equilibrium point at $\chi =0$,
the non-sphericity of the potential is essential for creation of the noticeable
angular momentum, i.e. for generation of baryonic charge asymmetry in the sector
of the scalar baryon $\chi$, see Fig.~\ref{fig:Chievol}. 

\begin{figure}[htbp]
	\centering
		\includegraphics[scale=0.7]{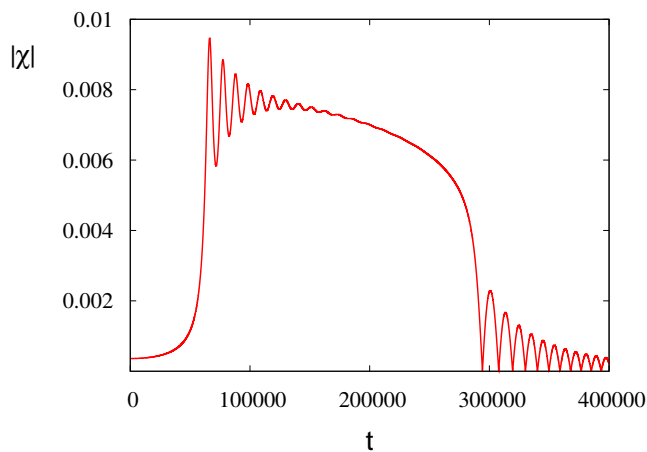}
		\caption{\small Evolution of $|\chi|$ in time. The field rolls down toward the 
		deeper minimum, 		oscillates there following the evolution of the minimum, 
		rolls back to the origin and starts to rotate around it.}
	\label{fig:Chimodevol}
\end{figure}

Thus to summarize. As it is already described above, during inflation,
when $\Phi$ is close to $\Phi_1$,
the quantum fluctuations of $\chi$ become large enough to overshoot the barrier
and field $\chi$ starts classical rolling down towards the deeper minimum, $\chi^{(2)}(t)$.
As one can see in Figs. \ref{fig:Chievol} and \ref{fig:Chimodevol}, $\chi$ reaches 
this minimum, follows its evolution in time, and oscillates around this minimum 
with decreasing amplitude. At the end of inflation
when the minimum of the potential at $\chi=0$ becomes again the only minimum,
the field rolls down to the origin along the slope of the potential. 
Coming close to the origin the field starts to rotate around it due to the angular momentum 
generated by non-sphericity of the potential, i.e. due to the complex mass terms (see 
Fig. \ref{fig:Chievol} and the zoomed part inside).  To gain a considerable angular
momentum the magnitude of this terms at low $\chi$
should be comparable with 
the others entering equation of motion (\ref{ddot-chi}). 
This is the reason why we have chosen $m_0$ only slightly larger than $|m_1|$. 
The term $\sim \chi^4$  is not essential near $\chi = 0$.  
The form and slope of the spiral-like trajectory depend on the 
effective mass of $\chi$. 
The rotation of $\chi$ around the origin corresponding to
the nonzero baryonic charge would be transformed into the baryonic asymmetry of 
quarks and ultimately of protons and neutrons by B-conserving decays of $\chi$
into (light or massless) fermions. The sign of the asymmetry depends on the
direction of the rotation and can be arbitrary because, as we mentioned above
the initial value of the phase $\theta$ is uniformly distributed in the interval
$[0,2\pi]$. So in this version of the scenario we would expect an equal number
of baryonic and antibaryonic bubbles. In slightly complicated models this symmetry
between B and anti-B bubbles may be lifted.

The magnitude of the asymmetry inside the B-bubbles can be
much larger than canonical value (\ref{beta}). Moreover, if the life-time
of $\chi$ noticeably exceeds the life-time of the inflaton, the asymmetry 
may be even larger than unity, because the energy density of the decay products of
the inflaton into relativistic particles would be strongly red-shifted with respect
to non-relativistic longer-lived $\chi$-bosons. The baryon asymmetry depends upon
the initial value of phase, $\theta$, of field $\chi$. According to the presented
above arguments, $\theta$ should be homogeneously distributed in the interval
$[0,2\pi]$. Thus the asymmetry between $\chi$ and anti-$\chi$ inside the  
bubbles may vary from zero to the maximum value $B_\chi \sim m |\chi|^2$.
The baryon asymmetry $\beta$ depends upon the ratio of 
the energy density of $\chi$, $\rho_\chi \sim m^2 |\chi|^2$, at the moment of $\chi$ 
decay to the background cosmological energy density. If $\chi$ decayed when the 
background relativistic matter was strongly red-shifted, the energy density inside
the bubble would be dominated by the energy density of $\chi$ and the ratio of the
baryonic charge density to the entropy density inside the bubbles would be about
\be
\beta_B \sim \frac{m |\chi|^2}{\left(m^2 |\chi|^2\right)^{3/4}} = 
\left(\frac{|\chi|}{m}\right)^{1/2}\pkt
\label{beta_B}
\ee 
This value may be much larger than unity.

\normalsize

\section{Distributions of bubbles over size and mass} \label{s-size}

The process of the B-bubble formation predominantly took place when the effective mass of 
$\chi$-field, eq.~(\ref{m-eff}), was smaller than the Hubble parameter (\ref{H2}). 
It roughly corresponds to
\be
\lambda_1 (\Phi - \Phi_1)^2 < \frac{4\pi m_\Phi^2 \Phi^2}{3m_{Pl}^2}\pkt
\label{m-eeff-H}
\ee
During this period we consider $\chi$ as a massless field and estimate how fast its quantum
fluctuations rose. Here we follow ref.~\cite{Dolgov:1992pu}, which in turn was based on
the diffusion equation governing evolution of the quantum fluctuations, derived
in paper~\cite{Starobinsky:1982ee}.
 
Qualitatively the picture looks as following. ``Massless'' field $\chi$ ``Brownian'' moves 
from zero with the average displacement given by eq.~(\ref{chi-2-vac}). If $\chi$
exceeds a certain value $\chi_c$, after which the motion of $\chi$ away from the origin 
would be governed by its potential (\ref{U-of-chi-Phi}), the motion turns into classical roll
down to the second minimum $\chi_2$  of this potential. The boundary value when the evolution
of $\chi$ turns from the quantum regime into the classical one can be found from the
condition $U''(\chi,\Phi) = H^2$, or approximately: 
\be
2\lambda_2 |\chi_c|^2 \ln \left(\frac{|\chi_c|^2}{\sigma^2} \right) \approx
\frac{4\pi m_\Phi^2 \Phi_1^2 }{3 m_{Pl}^2}\pkt
\label{chi-c}
\ee
Those $\chi$'s which were able to reach this value would give birth to B-bubbles. 
Earlier formed B-bubbles would have considerably 
larger sizes because their creation took place at inflationary period and the 
ratio of the radii of two bubbles crated at $t_1$ and $t_2$ would be:
\be
r_1/r_2 = \exp\left(\int_{t_1}^{t_2} H dt\right)\pkt
\label{r1-r2}
\ee

Let us introduce the probability distribution $\calp(\chi,t)$ of quantum 
fluctuations to have the value $\chi=\chi_1+i\chi_2$ at time $t$. We 
assume that it is normalized to unity:
\be
\int_{-\infty}^{+\infty} d\chi_1 d\chi_2 \calp(\chi,t) = 1\pkt
\label{norm}
\ee

The probability of B-bubble formation per unit time and per unit volume can be
parametrized as:
\be
\frac{dW}{dt dV} =\int d\chi_1d\chi_2 \mu^4 \calp(\chi,t)\kma
\label{dW-dt-dV}
\ee
where $\mu$ is a parameter with dimension of mass. It is natural to expect 
that $\mu =H$. There are simply no other parameter in the massless limit.  
At inflationary stage we may assume that $H \approx const$. The initial
size of a B-bubble should be of the order of $r_{in} \sim 1/H$. Hence the fraction
of the total volume occupied by the bubbles is
\be
\nu(t) = \int^t_{t_{in}} dt'H\int_{\chi_{c}}^\infty d\chi_1d\chi_2\calp(\chi,t')\kma
\label{epsilon-of-t}
\ee
where we may formally take $t_{in} = -\infty$ but in reality it is determined
by the condition $m_{eff}\leq H$. To obtain the above result we took into account that the 
bubbles formed at time $t'$ exponentially expand together with the rest of the
universe and their radius rises as $r(t) = H^{-1} a(t)/a(t') $, where $a(t)$ is the 
cosmological scale factor. So after formation their volume fraction remains constant.  

Correspondingly,  we find that the volume distribution of B-bubbles, or better to 
say, their number density as a function of their size would have the form:
\be
N (r,t) \sim \int_{\chi_{c}}^\infty d\chi_1d\chi_2 \calp\left (\chi, t'(r,t)\right)\kma
\label{N-of-r-t}  
\ee
where $t'(r,t)$ can be found from the equation: $a(t)/a(t') = rH$ and $H$ is the
Hubble parameter during inflation. In particular,
if we remain at inflationary stage, then $a(t) = \exp (Ht)$ and 
$t'(r,t) =  t -  H^{-1}\ln (rH)$. This derivation is valid for constant
$H$. In realistic situation $H$ drops down with time. We will consider this
case below.

As is argued in ref.~\cite{Dolgov:1992pu} (see also this section below),
for the quadratic potential $U = m^2_{eff}(t) |\chi|^2$
the probability distribution has the Gaussian form:
\be
\calp\left (\chi, t\right) = C(t) \exp \left[ -|\chi^2|/2 \langle \chi^2\rangle
\right]\kma
\label{gauss}
\ee
where $C(t)$ is determined by the normalization condition (\ref{norm}). The
mean square width of the distribution is given by
\be
\langle \chi^2\rangle = \frac{H^3}{4\pi^2}\int_{t_{i}}^t dt' 
\exp \left(-\frac{2}{3H}\int_{t'}^t dt'' m^2_{eff} (t'')\right)\pkt
\label{chi2-avrgd}
\ee 
When $\Phi$ is close to $\Phi_1$, the effective mass behaves as 
$m_{eff}^2 (t) \approx m_0^2 + m_3^4 (t-t_1)^2$, where $t_1$ is the moment
when $\Phi =\Phi_1$. This case was analyzed in ref.~\cite{Dolgov:1992pu},
where it was shown that both the bubble distributions over length and mass
have log-normal form. In particular,
\be
\frac{dn}{dM} = C_M \exp\left[-\gamma\,\ln^2(M/M_0) \right]\kma
\label{dn-dM}
\ee
where $C_M$, $\gamma$, and $M_0$ are some constant parameters, which 
can be expressed through the parameters of the model. 
For phenomenological applications we take $C_M$, $\gamma$, and $M_0$ as almost free
parameters with the values which are in the interval dictated by the ``reasonable''
values of the coupling constants and masses of $\chi$ and $\Phi$.
Notice that a modification of this distribution by a power factor, $M^b$, or,
which is the same, by a log term in the exponent, $\exp(b\ln M)$ lead to the
same log-normal form of the distribution (\ref{dn-dM}) with the corresponding 
change of the values of the parameters.
 
The log-normal form of the bubble distribution was derived in ref.~\cite{Dolgov:1992pu}
under simlifying assumption of a constant Hubble parameter during inflation.
In many realistic situations $H$ depends upon time and drops down practically to 
zero when inflation ends at $t=t_e$: 
\be
H(t) \equiv H_1 h(\tau) = H_1 [1 - h_1(\tau - \tau_1)].
\label{h-of-t}
\ee
Here $\tau = H_1 t$ and $t_1$ is the moment when $\Phi = \Phi_1$
and $h_1 = 1/(\tau_e - \tau_1) \ll 1$. 
For such $H(t)$ the size of the bubble created at $t=t_p$ rises as 
\be
l(\tau,\tau_p) = l_{in} \exp \left[\int_{\tau_p}^{\tau} d\tau h (\tau) \right]\pkt
\label{l-t-tp}
\ee
Correspondingly, the size of the bubble at the end of inflation will be
\be
l(\tau_e,\tau_p) = l_{in} \exp\left[\left(\tau_e-\tau_p\right)\left(1 + h_1\tau_1 -
\frac{1}{2} h_1\left(\tau_e+\tau_p\right)\right)\right]\pkt
\label{l-of-te-tp}
\ee
From this equation we see that the bubbe size is related to the production 
moment as
\be
\tau_p = h_1^{-1} \left( 1 +h_1 \tau_1 - \sqrt{2h_1 \ln l H_1}\right)\pkt
\label{t-p}
\ee 
The distribution of the bubbles over their size can be written:
\be
W(l) = C_l \exp\left[ -\gamma\left(\tau_e -\tau_1 -\sqrt{2 (\ln l)/h_1}\right)^2\right]\kma
\ee
After some modifications this expression can be rewritten as
\begin{eqnarray}
W(l) = C_l \exp\left[ -\gamma(\tau_e -\tau_1)^2\left(1
-\sqrt{\frac{2}{h_1(\tau_e-\tau_1)^2}\ln l}\right)^2\right]\pkt
\end{eqnarray}
Now after simple algebra we find that the distribution over $l$ again has the
log-normal form:
\be
W(l)&=&C_l \exp\left[ -\gamma(\tau_e -\tau_1)^2\left(1 -\sqrt{1+\frac{\ln l/l_0}{\ln l_0}}\right)^2\right]
\nonumber\\&\simeq& C_l \exp \left[ -\tilde{\gamma} \ln^2 \left(l/l_0\right)\right]\kma
\label{W-of-l}
\ee
where $l_0 = \exp[h_1(\tau_e-\tau_1)^2/2]$ and $\tilde{\gamma}=\gamma/h_1^2(\tau_e-\tau_1)^2$.

The equation governing the evolution of the quantum fluctuations of a real scalar
field was derived in ref.~\cite{Starobinsky:1982ee}. Following this derivation we can 
conclude that the probability distribution for a complex field $\chi$ in potential 
$U(\chi)$ in (quasi) De Sitter space-time with the Hubble parameter $H$ satisfies 
practically the same equation
\begin{equation}
\frac{\partial\calp}{\partial t}=
\frac{H^3}{8\pi^2}\sum_{k=1,2}\frac{\partial^2\calp}{\partial\chi_k^2}
+\frac{1}{3H}\sum_{k=1,2}\frac{\partial}{\partial\chi_k}\left[\calp\frac{\partial U}
{\partial\chi_k}\right]\kma
\label{dP-dt}
\end{equation}
where $\chi_{1,2}$ are the real and imaginary parts of field,
$\chi= \chi_1+i\chi_2$, respectively.

We decompose $\chi_k$ into classical part $\chi^{(0)}_k$ and quantum
fluctuations, 
\be
\chi_k = \chi^{(0)}_k + \eta_k\pkt
\label{chi-0-eta}
\ee
At the initial stage, when $\Phi$ was far from $\Phi_1$, the effective mass of
$\chi$ was large and quantum fluctuations were small near $\chi^{(0)}_k =0$. In this
case the potential governing their evolution can be approximated as the
quadratic one:
\begin{equation}
U_{q}=\frac{1}{2}m_{ij}^2\eta_i\eta_j\kma
\label{U-q}
\end{equation}
with the effective mass:
\begin{equation}
m_{ij}^2=\left.\frac{\partial^2U}{\partial\chi_i
\partial\chi_j}\right|_{\chi_i=\chi_i^{(0)}}\pkt
\label{m-ij}
\end{equation}
Due to the last two terms in eq. (\ref{U-of-chi-Phi}) the potential is not symmetric
with respect to the phase rotation of $\chi$ and the matrix $m_{ij}$ may contain
off-diagonal terms, $m_{12}\neq 0$. However by a suitable redefinition of 
$\chi\rightarrow \exp(i\phi)\chi$ we can eliminate imaginary part of $m$ and
make $m_{ij}$ diagonal.

For a quadratic potential eq.~(\ref{dP-dt}) can be solved in the form:
\begin{equation}
\calp=\cala(t)\exp(-a_{ij}\eta_i\eta_j)\kma
\label{P-quad}
\end{equation}
where $a_{ij}$ is a symmetric $2\times 2$ matrix and $\cala$ is
determined by the normalization condition (\ref{norm}):
\begin{equation}
\cala(t)=\frac{1}{\pi}\sqrt{\det{a}}\pkt
\label{A-of-t}
\end{equation}
After simple algebra, comparing terms of zeroth and second order in $\chi_i \chi_j$,
we obtain in the matrix form:
\be
\frac{1}{2}\tr [a^{-1} \dot a] &=& 
-\frac{H^3}{4\pi^2} \tr [a] + \frac{1}{3H} \tr [m^2]\kma
\label{tr-dot-a}\\
\dot a &=& -\frac{H^3}{2\pi^2}\,a^2 + \frac{2}{3H}\, a m^2\pkt
\label{dot-a} 
\ee
There are three unknown functions, $a_{ij} (t)$ and four equations (taken into
account symmetry of matrix $a$). Still the system is not over-determined. If we
multiply the second equation by matrix $a^{-1}$ and take trace, we obtain the
first equation.

\begin{figure}[htbp]
	\centering
	\hspace{-3cm}
		\includegraphics[scale=0.8]{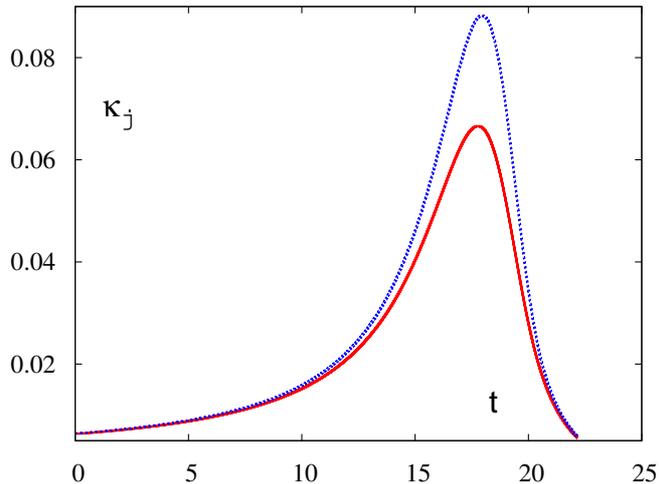}
		\caption{\small The variation of the mean square values
$\kappa_j$ as a function of time. All quantities are normalized to $H_1$,
i.e. $t\rightarrow H_1 t$ and $\kappa_j\rightarrow \kappa_j/ H_1^2$. This 
result is obtained with effective mass (\ref{m-ij}) using potential
		 (\ref{U-of-chi-Phi}).}
	\label{fig:kappa}
\end{figure}

As we mentioned above, the mass matrix $m^2$ can be made diagonal by phase
rotation of $\chi$. Correspondingly matrix $a$ becomes diagonal too and
the equations for $a_{11}$ and $a_{22}$ components decouple. Introducing
$\kappa_j = 1/a^{-1}_{jj}$ (no summation), we obtain the linear equations:
\be
\dot \kappa_j + \frac{2}{3H}\,m_j^2 \kappa_j = \frac{H^3}{2\pi^2}\pkt
\label{dot-kappa}
\ee
These equations can be integrated by quadratures for arbitrary $m^2(t)$
and $H(t)$ but it may be simpler to integrate them numerically directly 
(see Fig. \ref{fig:kappa}). 

\normalsize

This result allows to calculate the probability distribution for sufficiently
small quantum part of $\chi$, when quadratic approximation is valid.
As we have already mentioned, when $\chi$ reached the critical value 
$\chi_c$ (\ref{chi-c}) its evolution would be governed by classical equation
(\ref{ddot-chi}) with the potential which is dominated by log-quartic
term $\lambda_2 |\chi|^4 \ln |\chi|^2/\sigma^2$. These terms in the potential
become essential rather early, still at the quantum regime. To get a qualitative 
understanding of the impact of this terms we proceed as follows. We approximate 
the potential by
\be
U\approx m_{eff}^2 |\chi|^2 - \lambda_2 |\chi|^4\kma
\label{U-appr}
\ee
where $m_{eff}^2 =m_0^2+\lambda_1 (\Phi - \Phi_1)^2$. We neglect the last two
aspherical terms in the potential (\ref{U-of-chi-Phi}) because their role 
at this stage is not important. For a spherically symmetric potential
$U=U(|\chi|)$ the probability distribution depends only upon $|\chi|$ and
we will look for the approximate solution in the form:
\be 
\calp (\chi,t) = \cala (t) \left[ 1 + b(t)|\chi|^4 \right]\, 
\exp \left[ - |\chi|^2 /\kappa (t)\right] \kma
\label{P-of-b}
\ee
assuming that terms containing $b(t)$ with higher powers of $\chi$ are small.

The normalization condition now leads to:
\be
\cala = \frac{1}{\pi\kappa (1 + 2 b \kappa^2)}\pkt
\label{norm-b}
\ee
Substituting this expression into eq.~(\ref{dP-dt}) and neglecting terms 
of higher order in $b$ and sixth power of $\chi$ we find for zero, second, and
fourth powers of $\chi$ respectively (only two out of these three equations are
independent):
\be
\frac{\dot\kappa}{\kappa} &=& \frac{H^3}{2\pi^2\kappa}\left(1 + 4b\kappa^2\right)
- \frac{4}{3H}\,\left(m_{eff}^2 +4\lambda_2\kappa\right)\kma
\label{dot-kappa-2}\\
\dot b &=& -\frac{2H^3 b}{\pi^2 \kappa} + \frac{8\lambda_2}{3H\kappa}+ 
\frac{8m^2_{eff}b}{3H}\pkt
\label{dot-b}
\ee
It is easy to solve these equations numerically. As expected, $\kappa$ becomes a few percent
larger at $\Phi\sim\Phi_1$, i.e. the additional negative quartic term stimulates
the growth of fluctuations.

\section{Density perturbations created by B-bubbles and their evolution}\label{s-perturb}

As it was mentioned in ref.~\cite{Dolgov:1992pu}, B-bubbles could create large isocurvature density
perturbations at relatively small scales. The spectrum of angular fluctuations of 
cosmic microwave background radiation (CMBR) excludes noticeable isocurvature perturbations
with wave lengths larger than approximately 10 Mpc~\cite{Hinshaw:2008kr,Komatsu:2008hk}, 
but at smaller scales they are allowed and may even be very large.

In the considered model there are two mechanisms of the density perturbations, which were operative at different
cosmological epochs. Inside B-bubbles the baryon asymmetry, $\beta$, may be much larger
than the observed value $\beta \approx 6\cdot 10^{-10} $. Such inhomogeneous distribution 
of the baryon number would not lead to noticeable density
perturbations till quarks remained ultra-relativistic, i.e. till the QCD phase transition.
In this way relatively large objects with high density contrast could be created. This was
considered in ref.~\cite{Dolgov:1992pu}, where stellar and larger mass objects were studied.
There is another possibility to perturb the cosmological
energy density disregarded in ref.~\cite{Dolgov:1992pu}. This mechanism 
operated at an earlier cosmological period and might lead to perturbations with 
much smaller wave lengths and masses, $M\ll M_{\odot}$, than those created at the 
QCD phase transition. Here we consider in detail both possibilities.

Let us first discuss the early stage when the B-bubbles were
created. Initially, at inflationary stage, the density contrast between low and 
high $\chi$-regions was negligible, first, because the energy density of $\chi$
is assumed to be strongly subdominant with respect to the inflaton one. Second, the
bubble formation did not lead to a density contrast at the moment of formation
simply because of energy conservation. The energy density contrast $\delta\rho$ 
between the bubbles and the rest of the universe remained small
till the end of inflation. After inflation is over and the inflaton decayed 
into (predominantly) relativistic particles, the situation might change, if the
life-time of $\chi$ is larger than the time of the inflaton decay. 
Outside the bubbles there would be only relativistic matter, while
inside the bubbles there would be a mixture of nonrelativistic and relativistic
matter. Indeed, in the course of the subsequent
cosmological evolution the equation of state of $\chi$ would be different for 
small $\chi$ near zero and for large $\chi$ staying in the second minimum of
$U(\chi,\Phi)$. At this stage the energy density of $\chi$ would behave as
vacuum-like energy. When this minimum disappeared, $\chi$ started to oscillate 
around zero in essentially quadratic potential. So its equation of state became that of
nonrelativistic matter and its energy density was red-shifted as the scale factor
cube. 

To illustrate the evolution of the density perturbations 
in the considered model let us assume
that there are two large (larger than horizon) pieces of the universe, the bulk with
the relativistic equation of state, $p_r =\rho_r/3$, and a  bubble with a large 
value of $\chi$ with a mixture of relativistic and a small fraction of 
non-relativistic matter, $\rho_{nr}^{in}=\epsilon\rho_r^{in}$. 
The energy density inside the bubble evolves as:
\begin{equation}
\rho_{inside}=\frac{\rho_r^{in}}{x^4}+\frac{\rho_{nr}^{in}}{x^3}
=\frac{\rho_r^{in}}{x^4}\left(1+\epsilon x\right)\kma
\label{rho-inside}
\end{equation}
where $x=a/a_{in}$ is the ratio of the running cosmological scale factor to its initial
value. 

We can find the dependence of the scale factor on time solving the Friedmann equation:
\be
\left(\frac{\dot x }{ x}\right)^2 = \frac{8\pi}{3 m_{Pl}^2}\,\left(
\frac{\rho_r^{in}}{x^4} +  \frac{\rho_{nr}^{in}}{x^3}\right)\pkt
\label{dot-x}
\ee
Technically simpler is to solve this equation in terms of conformal time,
$d\tau = dt/a(t)$:
\be
{a} = a_{eq} \left(2 \eta +\eta^2 \right)\kma
\label{x-of-eta}
\ee
where $a_{eq} = a_{in}/\epsilon$ is the value of the scale factor when the energy
densities of relativistic and nonrelativistic matter inside the bubble
become equal and $\eta = \tau a_{in}/(4 t_{in} \epsilon)$. At 
matter radiation equilibrium  $\eta \sim 1$.
 
In what follows we will use the results presented in book~\cite{Mukhanov:2005sc}
where a very similar case of evolution of density perturbations in the mixed
relativistic and nonrelativistic matter is considered.
The gravitational potential $\Phi$ satisfies eq. (7.49) of this book
with the right hand side which is non-zero because of the difference of equations
of state inside and outside the bubble. The solution for $\Phi$ with the initial
condition $\Phi (\eta = 0) = 0$ has the form:
\be 
\Phi = \frac{\eta}{5} \frac{\eta^2 + 6\eta + 10}{(\eta+ 2)^3}\, 
\frac{\delta S}{S} \kma
\label{Phi}
\ee
where 
\be
\frac{\delta S}{S} = \frac{3}{4}\,\frac{\delta\rho_r}{\rho_r} -
\frac{\delta\rho_{nr}}{\rho_{nr}}\pkt
\label{delta-S}
\ee
Notice that the entropy perturbations are not small becasue 
$\delta\rho_{nr} /\rho_{nr} = 1$. 

The density perturbations are related to the potential through the equation
(7.47) of book~\cite{Mukhanov:2005sc}:
\be
\Delta \Phi - 3{\cal H} \left( \Phi' +{\cal{H}} \Phi\right) =
4\pi G_N a^2 \delta \rho,
\label{Delta-Phi}
\ee
where the Laplacian term in the l.h.s. can be neglected for large wave length
of perturbations. Thus $\delta\rho/\rho\simeq 2\Phi$.

From this equation we find that asymptotically at MD stage, 
i.e. for $\eta\gg 1$, the density perturbations tend to a constant value
\be
\left.\frac{\delta \rho}{  \rho}\right|_{MD}  = - \frac{2}{5}\, \frac{\delta S} {S} 
\approx \frac{2}{5}\pkt
\label{delta-rho-MD}
\ee
For small $\eta$, i.e. at RD stage the perturbations rise linerly with
conformal time:
\be
\left.\frac{\delta \rho}{  \rho}\right|_{RD}  = - \frac{\eta}{2} \frac{\delta S} {S} 
\approx \frac{\eta}{2}\pkt
\label{delta-rho-RD}
\ee

At some moment field $\chi$ should decay into relativistic matter and after that
both inside and outside the bubble equations of state would be identical,
those of relativistic matter, $p=\rho/3$. After decay the developed density 
perturbations would stay constant, see eq. (7.59) of ref.~\cite{Mukhanov:2005sc}.
In the instant decay approximation the initial value of the constant potential
and density contrast can be read off equations (\ref{delta-rho-MD}) and 
(\ref{delta-rho-RD}). To be more accurate we can proceed as follows. 
We have to solve 
the system of the following two equations, assuming that all nonrelativistic 
matter is created by $\chi$-field: 
\begin {eqnarray}
&&\dot{\rho}_\chi + 3H\rho_\chi + \Gamma_\chi\rho_\chi=0\label{dot-rho-chi}\kma\\
&&\dot{\rho}_r+4H\rho_r- \Gamma_\chi\rho_\chi=0\kma
\label{dot-rho-r}
\end{eqnarray}
where $\Gamma_\chi=1/\tau_\chi$ is decay rate of field $\chi$ into light particles.

The solution of equation (\ref{dot-rho-chi}):
\begin{equation}
\rho_\chi=\frac{\rho_\chi^{in}}{x^3}\exp{(-\Gamma_\chi t)}\pkt
\end{equation}
Equation (\ref{dot-rho-r}) would be simplified in terms of the new function
$\rho_r = \rho_r^{in} \zeta/x^4$ with the initial value $\zeta = 1$ at $t=0$, or
$x=1$. The resulting equation takes the form:
\begin{equation}
\dot{\zeta}=\epsilon\Gamma_\chi\,x e^{-\Gamma_\chi t}\kma
\label{dot-zeta}
\end{equation}
where $x=x(t)$ is determined by eq. (\ref{dot-x}) but instead of $\rho$ given by 
eq. (\ref{rho-inside}) we have to substitute there the solutions of 
eqs. (\ref{dot-rho-chi},\ref{dot-rho-r}). A solution to this system of equations
can be done numerically. 

We see that the density contrast for superhorizon perturbations initially grows up
but asymptotically tends to a constant value.
When the bubble size becomes smaller than cosmological horizon, the further
evolution of the bubble
depends on the magnitude of the density contrast at the moment of the horizon 
crossing, $\Delta_h$. If $\Delta_h \sim 1$, such bubbles would form primordial
black holes (PBH). For a review of PBH creation see e.g. 
refs.~\cite{Carr:2003bj,Dokuchaev:2007mf}. 

We should remember however, that the presented above estimates are true only
for small perturbations when first order expansion in perturbed quantities is
accurate. As we saw, the estimated perturbations are of the order of unity and
so we can trust the calculations only by an order of magnitude.

The mass inside the horizon is 
\be
M_h \sim m_{Pl}^2 t\pkt
\label{M-h}
\ee
So the discussed mechanism allows for formation of PBH with the masses of the order
of $M_{BH} \sim 4\cdot10^{38}(\tau_\chi/{\rm sec}){\rm g}$, if the size of the bubble
was inflated up to horizon at that moment. An oversimplified estimate of the life-time
of $\chi$-boson as $\tau_\chi \sim (g^2 m_\chi/4\pi)^{-1}$, 
where $g$ is the Yukawa coupling 
constant of $\chi$ to fermions, gives for $g^2=0.01$ and $m_\chi = 10^8$ GeV:
$M_{BH} \sim 10^{9} $ g. Such light PBHs should evaporate in about one second and 
would not be important in the present day universe. On the other hand, the life-time
estimated above is certainly too short.  In supersymmetric theory with conserved
$\cal R$-parity the lightest supersymmetric particle (LSP) must be stable. So $\chi$
should either be stable or decay into a massive LSP. For simplicity we will not dwell 
on the problems related to color conservation and formation of $\chi$ condensate. This
discussion can be found e.g. in the reviews~\cite{Barbier:2004ez,Kazakov:1997xb}. We 
will assume that
$\cal R$-parity is broken and $\chi$ predominantly decays into ordinary particles but 
with baryonic charge conservation. If this is realized, the life-time may be much 
smaller than the estimate presented above and PBHs with much larger masses could be
created. Since $\cal R$-parity violating effects are only bounded from above, the
life-time is allowed to be arbitrary large. Another effect which might result in 
larger PBH masses is a delayed evaporation of the $\chi$-condensate found in
refs.~\cite{Dolgov:1989sy,Dolgov:2005nf}. 
On the other hand, it is not obligatory to live in a SUSY world but instead we
can postulate {\it ad hoc} an existence of a new $\chi$-field with the necessary properties
and in this case we would be free from the SUSY restrictions.
So, one way or another, the considered mechanism can lead to
formation of PBH which live long enough to be present in the contemporary universe.

If the density contrast is not sufficient for creation of PBH, the bubble would end up
as a more dense piece of relativsitic matter inside the relativistic cosmological 
backgound. The density contrast $\delta$ inside horizon is known to remains constant 
or to be more precise, may rise logarithmically (see e.g. ~\cite{Kolb:1990vq, Mukhanov:2005sc, Straumann:2005mz}),  
while outside horizon it drops down as $1/t$, as is mentioned above. 
The bubbles might be dissolved by diffusion but in relativstic regime the mean free
path is about $l_{free} \sim 1/(\alpha^2 T)$ and thus remains small in comparizon
with the bubble size which rises with the cosmological scale factor $a\sim 1/T$. So if
the bubble size was initially larger than $l_{free}$ it remains such during all 
relativistic stage. 

The second period when the perturbation in baryonic number density could be 
transformed into the energy density perturbation, happened after the QCD phase 
transition (p.t.) from deconfinement to confinement phase. After this p.t. practically
massless quarks turned into heavy baryons and the baryonic number 
fluctuations would be transformed into the energy density perturbations. 
If the baryon asymmerty inside the bubble has the value $\beta_B$ the density
contrast after the QCD p.t. would be
\be
\Delta=\frac{\delta\rho}{\rho} = \frac{\beta n_\gamma m_p}{(\pi^2/30) g_* T^4}
\approx 0.07 \beta_B\,\frac{ m_p}{T}\kma
\label{Delta-QCD}
\ee
where temperature $T$ should be smaller than the temperature of the QCD p.t., 
$T_{QCD}$. The latter is rather poorly known and according to 
different estimates it varies between 100-200 MeV. The quoted
results, however, were obtained for zero or negligible baryonic chemical 
potential, $\mu_B$. In our case $\mu_B$ could be large, but not large enough to 
shift the value of $T_{QCD}$ for more then $2\%$ \cite{Muller:1997tm, Lombardo}.

The bubble with $\Delta$ (\ref{Delta-QCD}) larger than unity at horizon crossing
would form PBH but now (in contrast to the case consided above) with very large
masses, starting from a few solar masses to possibly millions of $M_\odot$, 
if the bubble size is so large that it enters horizon at, say, $t= 10-100$ s, 
see eq. (\ref{M-h}) with $t\geq 10^{-5}$ s. 

If at horizon crossing $\Delta_B <1$, the bubble evoluton is detemermined by the
relation between its size, $l_B$ and the Jeans wave length:
\be
\lambda_J = c_s \left({\frac{\pi m_{Pl}^2}{\rho}}\right)^{1/2} \kma
\label{lambda-J}
\ee
where $\rho$ is the background energy density and
$c_s$ is the speed of sound. For relativistic plasma $c_s = 1/\sqrt{3}$
and the Jeans wave length is close to horizon $l_h =2t$. 

The Jeans mass is defined as
\be 
M_J=\frac{4\pi}{3}\rho {\lambda_J}^3\pkt
\label{M-J}
\ee
As is well known,
the regions with $M>M_J$ or with $\lambda >\lambda_J$ would decouple from the cosmological 
expansion and form gravitationally binded systems. In the regions with smaller mass
gravitational instability is not developed and the perturbations oscillate.

In the usual cosmological plasma the Jeans mass and wave length are very large 
before hydrogen recombination because of very small baryon asymmetry or a very large 
entropy per baryon, see e.g. book~\cite{Weinberg:1972}. After $e^+e^-$ 
annihilation, i.e. at $T<0.1 $ MeV the baryonic Jeans mass was 
about $10^4 M_\odot$ and rose up to 
$10^{18} M_\odot$ till hydrogen recombination. After recombination it sharply drops down to 
$10^6 M_\odot$ and continue to decrease as $a(t)^{-3/2}$.

In our case the entropy per baryon is not huge and the picture is quite different.
The energy density inside B-bubble, assuming that baryons are nonrelativstic, is
\be
\rho_B = n_B m_p + (\pi^2 /30) g_* T^4\kma
\label{rho-B}
\ee
where $n_B$ is the baryon nuber density and
$g_*$ is the efective number of the relativistic species. For photons $g_* =2$ and
for equilibrium photons and $e^\pm$ pairs $g_* = 5.5$. 

The pressure density is equal to:
\be 
P_B = (\pi^2 /90) g_* T^4 + n_B T\pkt
\label{P-B}
\ee

For adiabatic perturbations the ratio of the baryon number density to the photon number
density remains constant if entropy release by $e^+e^-$-annihilation is neglected.
Disregarding this and similar efects we obtain by an order of magnitude:
\be
c_s^2 = \frac{1}{3}\,\frac{T}{T+ m_p\beta }\pkt
\label{c-s-2}
\ee 
Correspondingly the Jeans mass at the QCD phase transition would be about $10 M_\odot$ 
if $\beta =1$. Assuming that the B-bubble is dominated by nonrelativsitc matter we find
that with further cosmological expansion $M_J$ would drop as $a^{3/2}$
because the energy density of baryons decreases as $a^{-3}$ and the temperature of
mixture of nonrelativistic baryons and photons decreases as $T\sim 1/a^2$. For smaller
$\beta$ the Jeans mass can be significantly larger.

Thus we see that after the QCD p.t. PBH with masses starting from a few solar masses 
and much higher can be created as well as gravitationally bound objects with masses
from a fraction of the solar mass up to thousands or more solar masses 
could emerge. These early formed objects would 
be made either from matter or antimatter with equal probability.

\section{Observational bounds and implications}\label{s-cosm-astro}

There are two features of the considered here model which may make it 
cosmologically and astrophysically interesting. 

First, according to the presented scenario all cosmological dark matter 
can be made out of baryons and antibaryons in the form of compact stellar type
objects or primordial black holes{\footnote {A different model of cosmological
dark matter made of baryons in the form of astronomically small quark 
lumps, $\sim 10^7$ g is suggested in ref.~\cite{Oaknin:2003uv}}}. 
These PBHs can be rather light, e.g
$10^{15}-10^{20}$ g or very heavy, up to millions solar masses. 
The number density of light PBHs is restricted by the condition that their
evaporation should not create too strong electromagnetic radiation to
exceed the observed one. The analysis, summarized in ref.~\cite{Carr:2003bj,Carr:2005zd}
excludes PBHs in a large interval of masses as the dominant part of cosmological
DM. However, the conclusion of ref.~\cite{Carr:2005zd} is valid only for a narrow
spectrum of PBH masses, when all PBHs have practically equal masses. In our
case the mass spectrum can be sufficiently wide to allow for light PBHs to be 
the dominant dark matter (DM) particles and to avoid condradiction with
the observed cosmic electromagnetic radiation. 

A possible manifestation of PBH with masses around $10^{16}-10^{17}$ g may be 0.511 MeV 
line observed in the Galactic bulge
~\cite{Johnson,Leventhal,Purcell,Milne:2001zs,Knodlseder:2005yq,Jean:2005af,Weidenspointner:2006nua}
and probably from the halo~\cite{Weidenspointner:2007rs}. 
As argued in ref.~\cite{Bambi:2008kx}
such PBHs may explain the observed flux and be abundant enough to make all
cosmological DM. As was observed recently the source of this line
is asymmetric with respect to the galactic center and is correlated with
the observed X-ray binaries~\cite{Weidenspointner:2008zz}. It make probable that the 
origin of the annihilated positrons are these low mass X-ray binaries,
though the estimated intensity of the line is about a half of the observed.
On the other hand, the distribution of PBHs may be also asymmetric and
correlated with the distribution of the binaries.

If the central mass in the distribution (\ref{dn-dM}) is close to the
solar mass, we would expect quite heavy cosmological dark matter ``particles''
made out of heavy PBH and dead or low luminocity stars with masses of
a fraction of solar mass and may be much larger. An analysis of the evolution
of such stellar like objects formed in the early universe will be made elsewhere.
As is argued in ref.~\cite{Dolgov:1992pu}, an interesting feature of
this scenario is that on the high mass tail of the distribution (\ref{dn-dM})
there may be superheavy PBH with masses of  millions solar masses. 
We can choose the parameters of the distibuiton such that the number of
the superheavy BH would be equal to the number of the observed large 
galaxies. So the model explains an early quasar creation which may seed
galaxy formation, as it was recently argued~\cite{DiMatteo:2007sq}. To make
a better qualitative statement one should study the amount of matter accretion
during life time of these superheavy BHs from the moment of their creation
when the universe was about 1 sec old till practically present time.

The observed in microlensing searches stellar mass low luminocity 
objects (MACHOS) could be such solar mass PBH or dead very early stars and 
antistars. According to ref~\cite{Tisserand:2006zx}, such objects may make at most
40\% of the total galactic DM.
However, again this bound is true only for narrow mass spectrum of MACHOS. If
the latter is sufficiently wide PBH can be safely put
below the upper bound curve and make 100\% of galactic DM. (A little here, 
a little there makes unity.) 

An argument against very heavy DM particles is presented in ref.~\cite{Metcalf:2006ms},
according to which the objects with masses above $10^{-2}M_\odot$ may
constitute not more than  88\% of cosmological DM at 95\% confidence level.
So we should either allow for 12\% to be in the 
form of relatively light PBHs or planet-like objects with masses below 
$2\cdot 10^{31}$ g or hope that 95\% CL is not strong enough to exclude
very heavy PBHs.

The second possibly testable feature of the model is a prediciton of
noticeable amount af antimatter, in the form of compact stellar-like objects.
Out of $\Omega_{DM} = 0.25$ a half may consist of antimatter. So we have almost
charge symmetric universe with the amount of matter 
$\Omega_B \approx 0.16$ and that of antimatter $\Omega_{\bar B} \approx 0.12$.
Of course the bulk of baryonic and antibaryonic matter can be inside PBHs and
thus unboservable (except for its gravity), however, it is not excluded that
a noticeable amlount of antimatter is in the form of evolved anti-stars or
antimatter clouds. There are quite strong bounds on the amount of
cosmic antimatter in the usual form. For example, the nearest antigalaxy
mast be at least at the distance of 10 Mpc~\cite{Steigman:1976ev}. In baryosymmeteric
cosmology the nearest domain of antimatter should be at the distance of (a few)
Gygaparsecs~\cite{Cohen:1997ac}. Bounds on the amount of galactic antimatter in 
``normal'' form in baryo-asymmetric cosmology are discussed in 
ref.~\cite{Belotsky:1998kz, Golubkov:2000xz, Fargion:2001yb}. However, in 
our case these bounds are not applicable, see ref.~\cite{Bambi:2007cc}, and one may 
expect that roughly a half of the galactic mass is made of antimatter in the 
form of compact low luminocity stars or PBH situated in the disk and in the
halo. Possible antistars in the galactic Bulge may be sources of the observed
positrons~\cite{Bambi:2007cc}.
 
In the regions with high (anti)baryonic number density primordial
nucleosynthesis (BBN) would lead to quite diferent amount of light elements.
There would be practically no deuterium, somewhat larger amount of $^4 He$
and $^7Li$, and anomallously larger abundances of heavier elements which
practically are not produced in the standard 
BBN~\cite{Matsuura:2004ss, Matsuura:2005rb, Matsuura:2007sb}. 
This feature can indicate that a cloud or a stellar type objects with anomalous light
element abundances may consist of antimatter.

According to ref.~\cite{Naselsky:2002,Kim:2008xs} the moment of recombination is sensitive to
the spatial distribution of baryonic matter and it might be  helpful for
restriction of the model parameters or its confirmation.

For very simple form of the interaction between $\chi$ and the inflaton 
field (\ref{U-int}) the log-normal mass spectrum of BBs has maximum at some
$M_0$ which  can be either close to normal stellar mass or much smaller, say,
$10^{15}-10^{20}$ g. However, with a minor modification of the discussed model
we can obtain a multi-maxima spectrum with several values of $M_0$. To achieve 
that we need to modify the interaction of $\chi$ with the inflaton as
\be
U_{int} = \lambda_k \Pi_j^k (\Phi -\Phi_j)^2/m_{Pl}^{(2k-1)}.
\label{U-int-2}
\ee  
Such an interaction may emerge from non-renormalizble Planck scale physics.

\section{Conclusions}\label{s-conclusions}

We reconsidered and further developed the scenario of ref.~\cite{Dolgov:1992pu}
and argued that a simple coupling of the Affleck-Dine scalar baryon to the inflaton
may lead to an efficient creation of PBHs either with relatively small masses,
$M_{BH}=10^{14}-10^{20}g$, or much larger, $M_{BH}\geq M_\odot$. The predicted 
log-normal mass spectrum of PBH (\ref{dn-dM}) very much differ from the power law one
which could be created by other mechanisms considered in the existing literature.
This difference is crucial for relaxing the observational bounds on PBH abundance.
The suggested mechanism may explain formation of all observed heavy black holes from 
millions to thousands or tens solar masses. There is no unique and compelling 
explanation in the existing literature. There is an interesting possibility that all 
cosmological DM is made of PBH or compact stellar mass objects (early formed and now 
dead stars) with log-normal mass spectrum. In particular, the observed MACHOS may be 
solar mass PBHs, or dead (anti)stars.

In the simplest version the mechanism considered here leads to significant creation 
of astronomical objects made of antimatter. Such objects should have anomalay abundances 
of (primordially formed) light elements. The search for cosmic antimatter 
\cite{Picozza}, if successed would be crucial for the conformation of the scenarion.

\section{Acknowlidgment}
We thank V. Mukhanov for the discussion on the evolution of the cosmological density 
perturbations and A. Drago for a kind conversation about the QCD 
phase transition.
MK thanks N. Suyama for helpful discussion.
The work is supported by Grant-in-Aid for Scientific Research from
the Ministry of Education, Science, Sports, and Culture (MEXT), Japan,
No. 14102004 (M.K.). This work was supported
by World Premier International Research Center Initiative° WPI Initiative), MEXT,
Japan.


\end{document}